\documentclass[superscriptaddress,prc,aps,twocolumn,showpacs]{revtex4}
\usepackage{amsmath,amssymb}
\usepackage{graphicx}
\newcommand{\Mainz}[1]
{\affiliation{Institut f\"ur Kernphysik, University of Mainz, D-55099 
Mainz,Germany}}

\newcommand{\Bonn}[1]
{\affiliation{Helmholtz-Institut f\"ur Strahlen- und Kernphysik, University of Bonn, D-53115 
Bonn, Germany}}

\newcommand{\Kent}[1]
{\affiliation{Kent State University, Kent, Ohio 44242-0001, USA}}

\newcommand{\Glasgow}[1]
{\affiliation{SUPA School of Physics and Astronomy, University of Glasgow, Glasgow G12 8QQ, 
United Kingdom}}

\newcommand{\Gatchina}[1]
{\affiliation{Petersburg Nuclear Physics Institute, 188300 Gatchina, Russia}}

\newcommand{\Dubna}[1]
{\affiliation{Joint Institute for Nuclear Research, 141980 Dubna, Russia}}

\newcommand{\Pavia}[1]
{\affiliation{INFN Sezione di Pavia, I-27100 Pavia, Italy}}

\newcommand{\GWU}[1]
{\affiliation{The George Washington University, Washington, DC 20052-0001, USA}}

\newcommand{\LPI}[1]
{\affiliation{Lebedev Physical Institute, 119991 Moscow, Russia}}

\newcommand{\Dalhousie}[1]
{\affiliation{Dalhousie University, Halifax, Nova Scotia B3H 4R2, Canada}}

\newcommand{\Halifax}[1]
{\affiliation{Saint Mary's University, Halifax, Nova Scotia B3H 3C3, Canada}}

\newcommand{\UniPavia}[1]
{\affiliation{Dipartimento di Fisica, Universit\`a di Pavia, I-27100 Pavia, Italy}}

\newcommand{\Basel}[1]
{\affiliation{Institut f\"ur Physik, University of Basel, CH-4056 Basel, Switzerland}}

\newcommand{\Tomsk}[1]
{\affiliation{Laboratory of Mathematical Physics, Tomsk Polytechnic University, 634034 
Tomsk, Russia}}

\newcommand{\Edinburgh}[1]
{\affiliation{School of Physics, University of Edinburgh, Edinburgh EH9 3JZ, United 
Kingdom}}

\newcommand{\INR}[1]
{\affiliation{Institute for Nuclear Research, 125047 Moscow, Russia}}

\newcommand{\Sackville}[1]
{\affiliation{Mount Allison University, Sackville, New Brunswick E4L 1E6, Canada}}

\newcommand{\Regina}[1]
{\affiliation{University of Regina, Regina, Saskatchewan S4S 0A2, Canada}}

\newcommand{\Zagreb}[1]
{\affiliation{Rudjer Boskovic Institute, HR-10000 Zagreb, Croatia}}

\newcommand{\ITEP}[1]
{\affiliation{Institute for Theoretical and Experimental Physics,
 SRC Kurchatov Institute, Moscow, 117218 Russia}}

\newcommand{\Amherst}[1]
{\affiliation{University of Massachusetts, Amherst, Massachusetts 01003, USA}}

\newcommand{\Bochum}[1]
{\affiliation{Institut f\"ur Experimentalphysik, Ruhr-Universit\"at,
 D-44780 Bochum, Germany}}

\newcommand{\UCLA}[1]
{\affiliation{University of California Los Angeles, Los Angeles, California 90095-1547, USA}}

\newcommand{\Jerusalem}[1]
{\affiliation{Racah Institute of Physics, Hebrew University of Jerusalem, Jerusalem 91904, Israel}}
 
\begin{document}

\title{Measurement of $\pi^0$ photoproduction on the proton at MAMI C}

\author{P.~Adlarson}\Mainz \\
\author{F.~Afzal}\Bonn \\
\author{C.~S.~Akondi}\Kent \\
\author{J.~R.~M.~Annand}\Glasgow  \\
\author{H.~J.~Arends}\Mainz \\
\author{Ya.~I.~Azimov}\Gatchina \\
\author{R.~Beck}\Bonn \\
\author{N.~Borisov}\Dubna \\
\author{A.~Braghieri}\Pavia \\
\author{W.~J.~Briscoe}\GWU \\
\author{S.~Cherepnya}\LPI \\
\author{F.~Cividini}\Mainz \\
\author{C.~Collicott}\Dalhousie \\ \Halifax \\
\author{S.~Costanza}\Pavia \\ \UniPavia \\
\author{A.~Denig}\Mainz \\
\author{E.~J.~Downie}\Mainz \\ \GWU \\
\author{M.~Dieterle}\Basel \\
\author{M.~I.~Ferretti Bondy}\Mainz \\
\author{L.~V.~Fil'kov}\LPI \\
\author{A.~Fix}\Tomsk \\
\author{S.~Gardner}\Glasgow \\
\author{S.~Garni}\Basel \\
\author{D.~I.~Glazier}\Glasgow \\ \Edinburgh \\
\author{D.~Glowa}\Edinburgh \\ %?
\author{W.~Gradl}\Mainz \\
\author{G.~Gurevich}\INR \\
\author{D.~J.~Hamilton}\Glasgow \\
\author{D.~Hornidge}\Sackville \\
\author{G.~M.~Huber}\Regina \\
\author{A.~K\"aser}\Basel\\
\author{V.~L.~Kashevarov}\Mainz \\ \LPI \\
\author{S.~Kay}\Edinburgh \\
\author{I.~Keshelashvili}\Basel\\
\author{R.~Kondratiev}\INR \\
\author{M.~Korolija}\Zagreb \\
\author{B.~Krusche}\Basel \\
\author{V.~V.~Kulikov}\ITEP \\
\author{A.~Lazarev}\Dubna \\
\author{J.~Linturi}\Mainz \\
\author{V.~Lisin}\LPI \\
\author{K.~Livingston}\Glasgow \\
\author{I.~J.~D.~MacGregor}\Glasgow \\
\author{D.~M.~Manley}\Kent \\ 
\author{P.~P.~Martel}\Mainz \\ \Amherst \\
\author{M.~Martemianov}\ITEP \\
\author{J.~C.~McGeorge}\Glasgow \\
\author{W.~Meyer}\Bochum \\
\author{D.~G.~Middleton}\Mainz \\ \Sackville \\
\author{R.~Miskimen}\Amherst \\
\author{A.~Mushkarenkov}\Pavia \\ \Amherst \\
\author{A.~Neganov}\Dubna \\
\author{A.~Neiser}\Mainz \\ 
\author{M.~Oberle}\Basel \\
\author{M.~Ostrick}\Mainz \\  
\author{P.~Ott}\Mainz \\   
\author{P.~B.~Otte}\Mainz \\
\author{B.~Oussena}\Mainz \\ \GWU \\
\author{D.~Paudyal}\Regina \\
\author{P.~Pedroni}\Pavia \\
\author{A.~Polonski}\INR \\  
\author{V.~V.~Polyanski}\LPI \\
\author{S.~Prakhov}\thanks{corresponding author, e-mail: prakhov@ucla.edu}\Mainz \\ \GWU \\ \UCLA \\
\author{A.~Rajabi}\Amherst \\
\author{G.~Reicherz}\Bochum \\   
\author{G.~Ron}\Jerusalem \\
\author{T.~Rostomyan}\Basel \\
\author{A.~Sarty}\Halifax \\  
\author{D.~M.~Schott}\GWU \\
\author{S.~Schumann}\Mainz \\
\author{C.~Sfienti}\Mainz \\
\author{V.~Sokhoyan}\Mainz \\ \GWU \\
\author{K.~Spieker}\Bonn \\
\author{O.~Steffen}\Mainz \\
\author{I.~I.~Strakovsky}\thanks{corresponding author, e-mail: igor@gwu.edu}\GWU \\
\author{Th.~Strub}\Basel \\
\author{I.~Supek}\Zagreb \\
\author{M.~F.~Taragin}\GWU \\
\author{A.~Thiel}\Bonn \\
\author{M.~Thiel}\Mainz \\
\author{L.~Tiator}\Mainz \\  
\author{A.~Thomas}\Mainz \\   
\author{M.~Unverzagt}\Mainz \\
\author{Yu.~A.~Usov}\Dubna \\
\author{S.~Wagner}\Mainz \\
\author{D.~P.~Watts}\Edinburgh \\
\author{D.~Werthm\"uller}\Glasgow \\ \Basel \\
\author{J.~Wettig}\Mainz \\
\author{L.~Witthauer}\Basel \\ 
\author{M.~Wolfes}\Mainz \\
\author{R.~L.~Workman}\GWU \\
\author{L.~A.~Zana}\Edinburgh \\

\collaboration{A2 Collaboration at MAMI}

\date{\today}

\begin{abstract}
Differential cross sections for the $\gamma p \to \pi^0 p$ reaction
have been measured with the A2 tagged-photon facilities at
the Mainz Microtron, MAMI C, up to the center-of-mass energy $W=1.9$~GeV.
The new results, obtained with a fine energy and angular binning,  
increase the existing quantity of $\pi^0$ photoproduction data 
by $\sim 47\%$. Owing to the unprecedented statistical accuracy 
and the full angular coverage, the results are sensitive to high
partial-wave amplitudes. This is demonstrated by the decomposition
of the differential cross sections in terms
of Legendre polynomials and by further comparison to model predictions.
A new solution of the SAID partial-wave analysis obtained after adding
the new data into the fit is presented. 
\end{abstract}


\maketitle

\section{Introduction}
\label{Intro}

 Measurements of pion photoproduction on both proton and
 quasifree-neutron targets have a very long history,
 started about 65 years ago.
 Using the first bremsstrahlung facilities, pioneering results for
 $\gamma p\to\pi^0 p$~\cite{pi0p}, $\gamma p\to\pi^+ n$~\cite{pi+n},
 and $\gamma n\to\pi^- p$~\cite{pi-p} were obtained.
 Despite all the shortcomings of the first measurements
 (such as large normalization uncertainties, wide energy
 and angular binning, limited angular coverage, etc.),
 those data were crucial for the discovery of
 the first excited nucleon state, $\Delta(1232)3/2^+$~\cite{Fermi}.

 Though present electromagnetic facilities in combination with
 modern experimental detectors have allowed a significant improvement
 in the quantity and quality of pion-photoproduction
 data~\cite{du13}, full understanding of pion-photoproduction
 dynamics is far from established, even in the best investigated
 area, comprising the first, second, and third resonance regions.
 The properties of many nucleon states in this energy
 range, especially in the third resonance region~\cite{PDG},
 are still not well understood.
 Resolving the complex partial-wave structure of pion photoproduction
 with minimal model assumptions requires more 
 precise data for differential cross sections, with fine energy
 binning and full angular coverage,
 in combination with measurements of various polarization observables.

 First results for polarization degrees of freedom have been recently
 reported by the A2, CBELSA/TAPS, and GRAAL Collaborations.
 The beam asymmetry $\Sigma$ has been measured by GRAAL~\cite{Bartalini_05}
 and CBELSA/TAPS~\cite{Thiel_12}. Data for the other two single polarization
 observables $T$ (target polarization) and $P$ (recoil-nucleon polarization
 measured as a double-polarization observable with transversely polarized
 target and linearly polarized beam) have been reported from
 CBELSA/TAPS~\cite{Hartmann_14}. First results for double-polarization asymmetries
 have also been measured at CBELSA/TAPS with a longitudinally polarized target
 (observable $G$ with linearly polarized beam~\cite{Thiel_12} and
  observable $E$ with circularly polarized beam~\cite{Gottschall_14})
 and with a transversely polarized target and linearly polarized beam
 (observable $H$~\cite{Hartmann_14}). The A2 Collaboration at MAMI obtained
 results for the double-polarization asymmetry $C_x^{\star}$ using a circularly
 polarized photon beam and a nucleon recoil polarimeter~\cite{Sikora_14}.
 Further data sets are currently under analysis, so that one can expect
 new analyses which provide much tighter constraints on
 the $\gamma p\to\pi^0 p$ reaction in the nearest future.

 This work contributes to these efforts by presenting a new high-statistics
 measurement of the $\gamma p\to\pi^0 p$ differential cross sections conducted
 by the A2 Collaboration for incident-photon energies,
 $E_\gamma$, from 218~MeV up to 1573~MeV [or center-of-mass (c.m.)
 energies $W=$1136--1957~MeV]. The data are obtained with a fine
 binning in $E_\gamma$ ($\sim 4$~MeV for all energies below
 $E_\gamma=1120$~MeV) and 30 angular bins, covering the full range
 of the $\pi^0$ production angle. The data obtained above
 $E_\gamma=1443$~MeV ($W=1894$~MeV) have limited angular coverage.

 A more detailed analysis of the present $\gamma p\to\pi^0 p$
 differential cross sections, combined with results from
 the A2 Collaboration obtained by measuring 
 polarization observables, is currently in progress and
 will be published separately.

\section{Experimental setup}
\label{sec:Setup}

The reaction $\gamma p\to \pi^0 p$ was measured using
the Crystal Ball (CB)~\cite{CB} as a central calorimeter
and TAPS~\cite{TAPS,TAPS2} as a forward calorimeter.
These detectors were installed at the energy-tagged
bremsstrahlung-photon beam produced from the electron beam
 of the Mainz Microtron (MAMI)~\cite{MAMI,MAMIC}.

The CB detector is a sphere consisting of 672
optically isolated NaI(Tl) crystals, shaped as
truncated triangular pyramids, which point toward
the center of the sphere. The crystals are arranged in two
hemispheres that cover 93\% of $4\pi$ sr, sitting
outside a central spherical cavity with a radius of
25~cm, which is designed to hold the target and inner
detectors.

 In the A2 experiments at MAMI C, TAPS was
initially arranged in a plane consisting of 384 BaF$_2$
counters of hexagonal cross section.
It was installed 1.5~m downstream of the CB center
and covered the full azimuthal range for polar angles
from $1^\circ$ to $20^\circ$.
In the present experiments, 18 BaF$_2$ crystals,
covering polar angles from $1^\circ$ to $5^\circ$, were replaced
with 72 PbWO$_4$ crystals. This allowed running with
a much higher MAMI electron current, without decreasing
the TAPS efficiency due to the very high count rate
in the crystals near the photon-beam line.
The energy resolution of the PbWO$_4$ crystals has not been
understood well yet, and the use of their information was 
restricted in the present analysis, as described later.  
More details on the calorimeters and their resolutions are given
in Ref.~\cite{slopemamic} and references therein.
\begin{figure*}
\includegraphics[width=0.85\textwidth,height=5.cm,bbllx=0.5cm,bblly=.35cm,bburx=20.cm,bbury=7.cm]{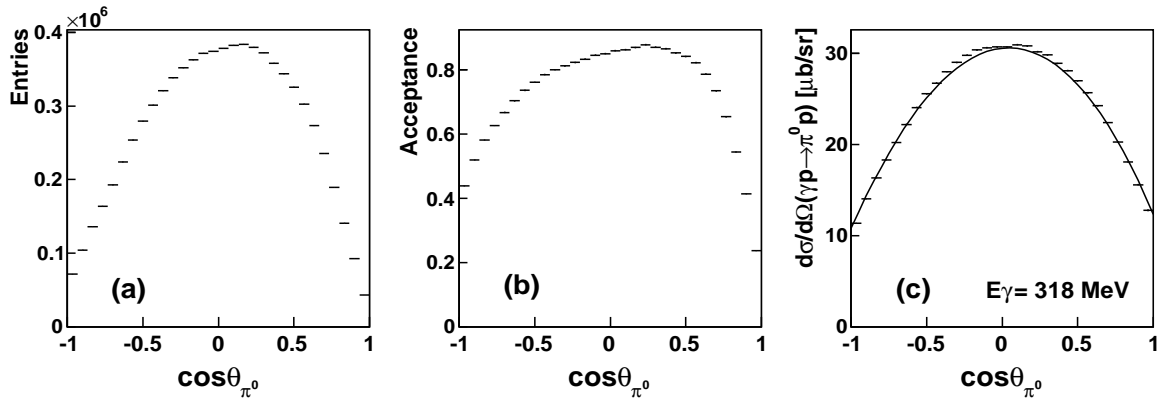}
\caption{
 $\cos\theta$ distributions for $\gamma p\to \pi^0 p$, with $\theta$ being
 the angle between the directions of the outgoing $\pi^0$ and the incident photon
 in the c.m.\ frame, obtained for $E_\gamma=318$~MeV ($W=1215$~MeV):
 (a)~experimental data after subtraction of the empty-target and
     the random backgrounds;
 (b)~angular acceptance obtained from the MC simulation of $\gamma p\to \pi^0p$;
 (c)~results for the $\gamma p\to \pi^0p$ differential cross section
     compared to the prediction by SAID~\protect\cite{SAID} with its
     CM12~\protect\cite{cm12} solution.
   The error bars on all data points represent statistical uncertainties only.
}
 \label{fig:dcs_pi0p_e318_apr13}
\end{figure*}
\begin{figure*}
\includegraphics[width=0.85\textwidth,height=5.cm,bbllx=0.5cm,bblly=.35cm,bburx=20.cm,bbury=7.cm]{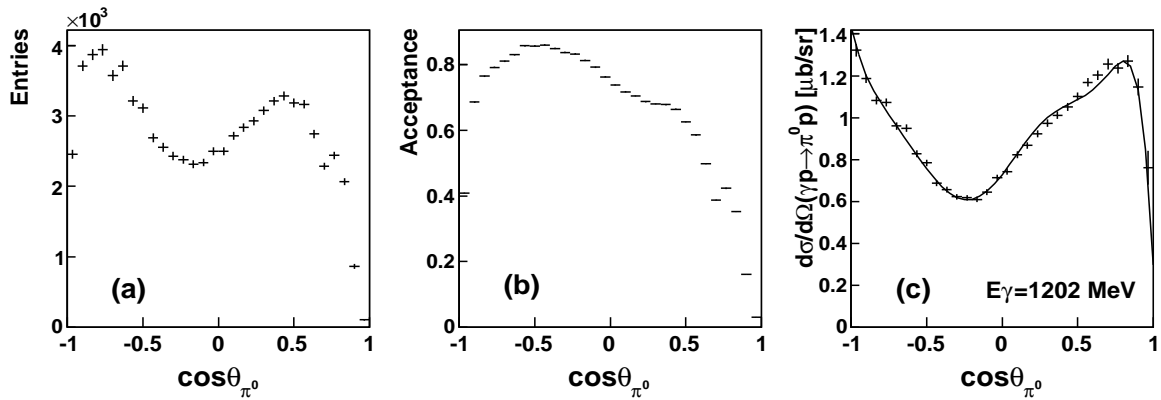}
\caption{
 Same as Fig.~\protect\ref{fig:dcs_pi0p_e318_apr13}, but for
 $E_\gamma=1202$~MeV ($W=1771$~MeV).
}
 \label{fig:dcs_pi0p_e1202_apr13}
\end{figure*}

 The present measurements used 1557-MeV and 1604-MeV
 electron beams from the Mainz Microtron, MAMI C~\cite{MAMIC}.
 The data with the 1557-MeV beam were taken in the first half of 2013
 (Run-I) and those with the 1604-MeV beam in the second
 half of 2014 (Run-II).
 Bremsstrahlung photons, produced by the beam electrons
 in a 10-$\mu$m Cu radiator and collimated by a 4-mm-diameter Pb collimator,
 were incident on a 10-cm-long liquid hydrogen (LH$_2$) target located
 in the center of the CB. In Run-I with the 1557-MeV electron
 beam, the energies of the incident photons were analyzed
 up to 1448~MeV by detecting the postbremsstrahlung electrons
 in the Glasgow tagged-photon spectrometer
 (Glasgow tagger)~\cite{TAGGER,TAGGER1,TAGGER2}.
 The uncertainty in the energy of the tagged photons is mainly determined
 by the number of tagger focal-plane detectors in combination with
 the energy of the MAMI electron beam used in the experiments.
 Increasing the MAMI energy results in increasing both the energy range covered
 by the spectrometer and the corresponding uncertainty in $E_\gamma$.
 For the MAMI energy of 1557~MeV and the Glasgow tagger, such an uncertainty was
 about $\pm 2$~MeV. The systematic uncertainty in the absolute value of $E_\gamma$,
 which is dominated by the energy calibration of
 the tagger, was about 0.5~MeV~\cite{TAGGER2}.
 More details on the tagger energy calibration and uncertainties
 in the energies can be found in Ref.~\cite{EtaMassA2}.

 In Run-II with the 1604-MeV electron beam,
 the energies of the incident photons were analyzed from
 1426~MeV up to 1576~MeV by detecting the postbremsstrahlung electrons
 in the Mainz end-point tagger (EPT) with 47 focal-plane detectors.
 Since the Glasgow tagger was not designed to
 measure the high-energy tail of the bremsstrahlung spectrum,
 the EPT spectrometer was built to conduct $\eta'$
 measurements by covering this low-energy range
 of postbremsstrahlung electrons.
 The uncertainty of the EPT in $E_\gamma$ due to the width of its
 focal-plane detectors was about $\pm 1.6$~MeV, with a similar value
 (i.e., $\sim 1.6$~MeV) in the systematic uncertainty in $E_\gamma$
 due to the EPT energy calibration. The energy calibration
 of the EPT is based only on the simulation of electron tracing,
 using measured magnetic-field maps. The correctness of
 this calibration, as well as its uncertainty, was checked by measuring
 the position of the $\eta'$ threshold, $E_\gamma \approx 1447$~MeV.
\begin{figure*}
\includegraphics[width=0.85\textwidth,height=5.cm,bbllx=0.5cm,bblly=.35cm,bburx=20.cm,bbury=7.cm]{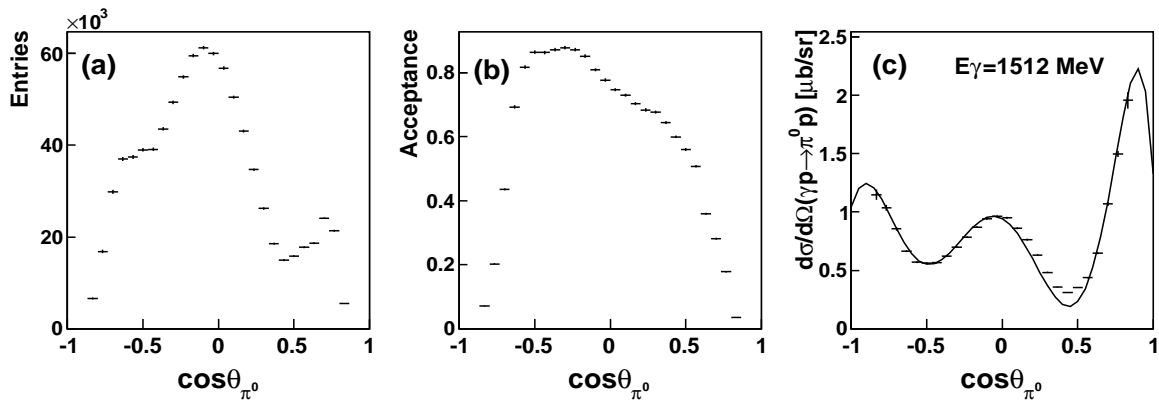}
\caption{
 Same as Fig.~\protect\ref{fig:dcs_pi0p_e318_apr13}, but for
 $E_\gamma=1512$~MeV ($W=1928$~MeV).
}
 \label{fig:dcs_pi0p_e1512_oct14}
\end{figure*}

 Because the main goal of Run-I
 was to measure the $\gamma p\to \pi^0 p$ reaction from
 $E_\gamma=218$~MeV up to 1448~MeV and with full coverage of the
 $\pi^0$ production angle, the trigger required the total energy deposited
 in the CB to exceed $\sim$120~MeV. Run-II was mainly
 dedicated to studying $\eta'$ physics. For that, the experimental
 trigger required the CB energy to exceed $\sim$540~MeV, rejecting
 the detection of $\pi^0$ production in the very forward
 and backward angles.

\section{Data handling}
\label{sec:Data}

 The reaction $\gamma p\to \pi^0 p\to \gamma\gamma p$ was searched for in
 events reconstructed with two or three clusters detected
 in the CB and TAPS together.
 In the analysis of the data from Run-I, the information from
 the 72 PbWO$_4$ crystals of TAPS was not used at all.
 In the analysis of the data from Run-II, in which all
 PbWO$_4$ crystals demonstrated good performance, their information
 was used in the cluster reconstruction.
 The two-cluster events were analyzed
 assuming that the final-state proton was not detected.
 This typically happens when the outgoing proton is stopped in the material
 of the downstream beam tunnel of the CB, or the proton kinetic energy
 in the laboratory system is below the software cluster threshold,
 which was 12 MeV in the present analysis.

 The selection of event candidates and the reconstruction of the reaction
 kinematics were based on the kinematic-fit technique.
 Details on the kinematic-fit parametrization of the detector
 information and resolutions are given in Ref.~\cite{slopemamic}.
 The kinematic-fit technique was also used for the offline calibration of
 the calorimeters. The resolution functions used in the kinematic fit were
 adjusted to result in proper stretch functions (or pulls) and
 probability distributions. All two- and three-cluster events
 that satisfied the $\gamma p\to \pi^0 p \to \gamma\gamma p$ hypothesis
 with a probability greater than 1\% were accepted for further analysis.
 The kinematic-fit output was used to reconstruct the four-momenta
 of the outgoing particles. The events from the data of Run-II
 with more than one cluster in the 72 PbWO$_4$ crystals of TAPS
 were discarded from the analysis, and the events with one cluster
 in these crystals were tested only for the hypothesis assuming that
 this cluster is produced by the recoil proton. Such a restriction
 was imposed due to the unknown energy-resolution function for
 the PbWO$_4$ crystals, and only the angular information
 from the proton cluster was used in the kinematic fit.

 The determination of the experimental acceptance was based on
 a Monte Carlo (MC) simulation of $\gamma p\to \pi^0 p\to \gamma\gamma p$
 with an isotropic production-angular distribution and a uniform
 beam distribution generated for the full energy range of the tagged
 bremsstrahlung photons.
 All MC events were propagated through a GEANT (3.21) simulation
 of the experimental setup. To reproduce the experimental resolutions,
 the GEANT output was subject to additional smearing, thus allowing
 both the simulated and experimental data to be analyzed in the same way.
 This additional smearing was adjusted by describing the experimental
 invariant-mass resolutions as well as the kinematic-fit probability
 distributions. The final adjustment of the detection efficiency for
 $\gamma p\to \pi^0 p\to \gamma\gamma p$ took into account the trigger
 requirements in the analysis of the MC events.

 MC simulations for other possible reactions
 (such as $\gamma p\to \eta p\to \gamma\gamma p$, $\gamma p\to \pi^0\pi^0 p$,
 $\gamma p\to \omega p \to \pi^0\gamma p$, $\gamma p\to \pi^0\pi^+ n$)
 showed that background contributions from these sources were negligibly small.
 Thus, experimental spectra with selected events were contaminated
 only by two sources of background events, the distributions of which
 were directly subtracted afterwards.
 The first source of background was due to interactions of
 incident photons in the windows of the target cell.
 The subtraction of this background was based on the
 analysis of data samples that were taken with an empty
 (no liquid hydrogen) target. The weight for the subtraction
 of the empty-target spectra was taken as a ratio of the photon-beam fluxes
 for the data samples with the full and the empty target.
 A second background was caused by random coincidences
 of the tagger counts with the experimental trigger;
 its subtraction was carried out by using
 event samples for which all coincidences were random
 (see Refs.~\cite{slopemamic,etamamic} for more details).

 For measuring the $\gamma p\to \pi^0 p$ differential cross sections,
 all selected events were divided into 30 equal-width
 $\cos\theta$ bins, covering the full range from --1 to 1,
 where $\theta$ is the angle between the directions of the outgoing $\pi^0$
 and the incident photon in the c.m.\ frame.
 Also, the quantity of data collected for $\gamma p\to \pi^0 p$ was sufficient
 to obtain statistically accurate differential cross sections for
 every tagger channel in the energy range from $E_\gamma=218$~MeV
 to 1120~MeV, providing $\sim 4$~MeV binning in $E_\gamma$.
 For higher energies, energy bins were combined in two or
 more tagger channels.

 The typical experimental statistics and acceptance are illustrated
 in Figs.~\ref{fig:dcs_pi0p_e318_apr13}, \ref{fig:dcs_pi0p_e1202_apr13},
 and \ref{fig:dcs_pi0p_e1512_oct14} for different energies and data sets.
 Figure~\ref{fig:dcs_pi0p_e318_apr13}(a) shows the experimental
 $\cos\theta$ distribution for $E_\gamma=318$~MeV ($W=1215$~MeV),
 which is obtained after subtracting the random and the empty-target
 backgrounds. Since this energy bin is close to the $\Delta(1232)3/2^+$
 maximum, the statistical uncertainties in this distribution,
 based on $8\times10^6$ events, are very small ($\sim 0.2\%$).
 Figure~\ref{fig:dcs_pi0p_e318_apr13}(b) shows the corresponding angular
 acceptance, which is greater than 80\% for the central range of $\pi^0$
 production angles. The resulting $\gamma p\to \pi^0 p$ differential
 cross section for $E_\gamma=318$~MeV is depicted in
 Fig.~\ref{fig:dcs_pi0p_e318_apr13}(c). It is also compared to
 the prediction from the partial-wave analysis (PWA) by SAID~\cite{SAID}
 with its CM12~\cite{cm12} solution (shown by a solid line).
 The experimental data points lie slightly
 above the phenomenological prediction, but their angular
 dependence is very close to it. Because all modern PWAs give
 very similar predictions for the $\gamma p\to \pi^0 p$
 differential cross section at the $\Delta(1232)3/2^+$ maximum,
 the present results in this region do not contradict any of them.
\begin{figure*}
\includegraphics[width=0.95\textwidth,height=5.cm,bbllx=0.5cm,bblly=.25cm,bburx=20.cm,bbury=5.7cm]{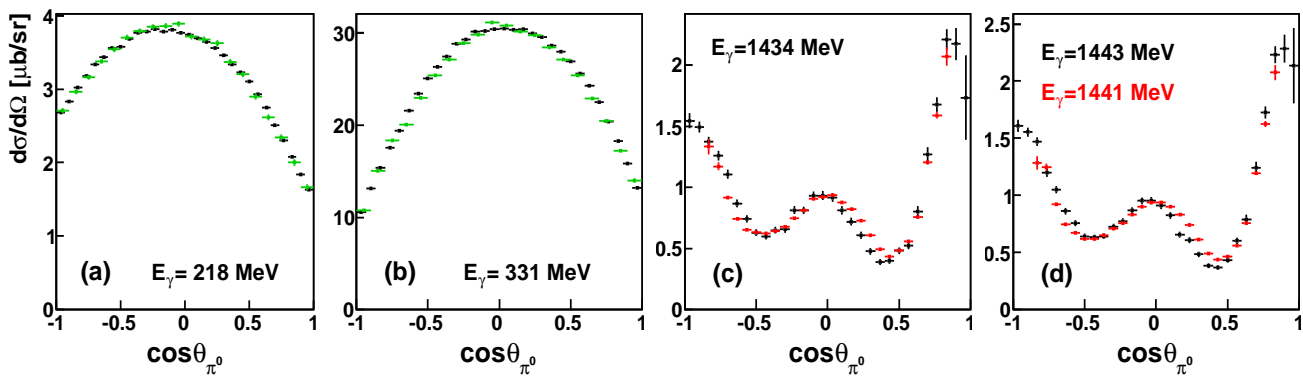}
\caption{(Color online)
 Comparison of the $\gamma p\to \pi^0 p$ differential
 cross sections obtained by the A2 Collaboration in different experiments:
 ~(a)~results for $E_\gamma=218$~MeV from Run-I
   (black points) and from the experiment~\protect\cite{hornidge13}
   with MAMI energy of 855~MeV (green points);
 ~(b)~same as (a) but for $E_\gamma=331$~MeV;
 ~(c)~results for $E_\gamma=1434$~MeV from Run-I
   (black points) and Run-II (red points);
 ~(d)~same as (c) but for $E_\gamma=1443$ and 1441~MeV of Run-I
   and Run-II, respectively;
  The error bars on all data points represent statistical uncertainties only.
 }
 \label{fig:dcs_pi0p_a2}
\end{figure*}

 Figure~\ref{fig:dcs_pi0p_e1202_apr13}(a) shows the experimental
 $\cos\theta$ distribution for $E_\gamma=1202$~MeV ($W=1771$~MeV), which is
 near the third resonance region. The statistical uncertainties in
 this distribution, based on $8.5\times10^4$ events, are larger but
 are still sufficient for a reliable analysis of the resulting
 differential cross section. Since the shape of the $\gamma p\to \pi^0 p$
 differential cross sections in this region changes
 rapidly with energy, the main goal of the present work was to
 measure it with the finest energy binning.
 Figure~\ref{fig:dcs_pi0p_e1202_apr13}(b) shows that the angular
 acceptance drops for the very forward angles. At these energies,
 this occurs as many $\pi^0$ mesons produced at forward angles
 have both their decay photons going into TAPS, and such events
 do not pass the trigger requirements.
 The resulting differential cross section, plotted in
 Fig.~\ref{fig:dcs_pi0p_e1202_apr13}(c), demonstrates again
 a quite reasonable agreement with the SAID CM12 solution~\cite{cm12}.

 Figure~\ref{fig:dcs_pi0p_e1512_oct14} shows the $\cos\theta$
 distributions for $E_\gamma=1512$~MeV ($W=1928$~MeV) from the data taken
 in Run-II. Despite the large quantity of the experimental events
 ($9\times10^5$), a very low, and even zero, acceptance for the very
 forward and backward angles, caused by the trigger, did not
 allow the first two and the last two $\cos\theta$ bins to be measured.
 However, the remaining 26 data points of the differential cross section
 lie close to the prediction from SAID CM12~\cite{cm12}.

\section{Experimental results and their discussion}
  \label{sec:Results}

 The $\gamma p\to \pi^0 p$ differential cross sections were obtained
 by taking into account the number of protons in the LH$_2$ target
 and the photon-beam flux from the tagging facilities corrected for
 the fraction rejected by collimators.
 The overall systematic uncertainty due to the calculation of
 the detection efficiency and the photon-beam flux
 was estimated to be 4\% for the data taken in Run-I
 and 5\% for the data taken in Run-II.

 Another source of systematic uncertainty is due to
 imperfections in reproducing the angular dependence of
 the $\cos\theta$ acceptance. Such imperfections
 are easier to see in differential cross sections with
 very low statistical uncertainties and a smooth shape.
 In the present data, it can be seen, for example,
 on the top of the experimental spectrum shown
 in Fig.~\ref{fig:dcs_pi0p_e318_apr13}(c).
 The angular dependence of differential cross sections
 can also be smeared due to the limited resolution in
 the $\pi^0$ production angle, resulting in a larger
 population of angular bins with a lower cross section.
 Such an effect is, possibly, seen in
 Fig.~\ref{fig:dcs_pi0p_e1512_oct14}(c) near $\cos\theta=0.5$.
\begin{figure*}
\includegraphics[width=0.95\textwidth,height=9.cm,bbllx=0.5cm,bblly=.25cm,bburx=20.cm,bbury=10.5cm]{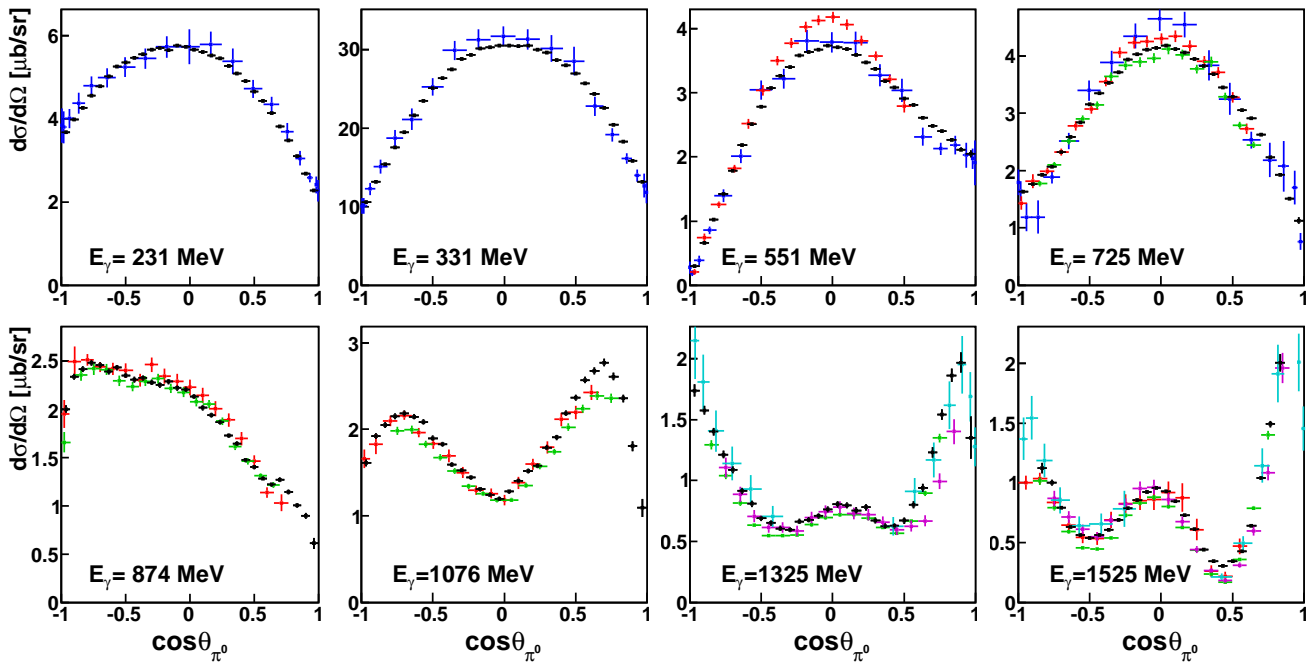}
\caption {(Color online)
 Comparison of the $\gamma p\to \pi^0 p$ differential
 cross sections from the present work (black points) with
 most recent results from other experimental setups at
 MAMI~\protect\cite{be06} (blue points)
 and other facilities:
 GRAAL~\protect\cite{Bartalini_05} (red points),
 CLAS~\protect\cite{du07} (green points),
 CB-ELSA~\protect\cite{ba05} (magenta points), and
 CBELSA/TAPS~\protect\cite{cr11} (cyan points).
 Values of $E_\gamma$ in each panel indicate the photon-beam
 energies for the results of this work. 
 All data from other experiments are shown for nearby energies.
 The error bars on the data points from this work represent statistical
 uncertainties only. The uncertainties of some previous results
 also include angular-dependent systematic uncertainties.
}
\label{fig:dcs_pi0p_a2_others}
\end{figure*}

 Possible magnitudes of systematic uncertainties in differential
 cross sections can be checked by comparing similar
 results obtained by analyzing different data. 
 Figure~\ref{fig:dcs_pi0p_a2} illustrates self-consistency
 of the results for the $\gamma p\to \pi^0 p$ differential
 cross sections obtained by the A2 Collaboration in three different
 experiments. The results for the lowest energy, $E_\gamma=218$~MeV,
 taken in Run-I are compared in Fig.~\ref{fig:dcs_pi0p_a2}(a)
 with similar results obtained in the A2 experiment (A2-2008) with
 the Glasgow tagger and MAMI energy of 855~MeV~\cite{hornidge13}.
 As seen, the agreement of the two independent measurements is fairly
 good for both the absolute values and the angular dependence.
 Figure~\ref{fig:dcs_pi0p_a2}(b) compares the results of Run-I
 and A2-2008 for an energy near the $\Delta(1232)3/2^+$ maximum,
 showing a small discrepancy in the angular dependence.
 Continuing the comparison of the results from Run-I and A2-2008
 at energies above the $\Delta(1232)3/2^+$ peak makes little sense
 as, at those energies, the A2-2008 data have much larger systematic
 uncertainties in the photon flux and in the angular acceptance,
 affected by the trigger.   
 The results for two the highest energies, $E_\gamma=1434$ and 1443~MeV,
 taken in Run-I are compared in Figs.~\ref{fig:dcs_pi0p_a2}(c) and
 (d) with similar results from the present work obtained in Run-II.
 Despite a general agreement, there are some discrepancies
 between the two sets of results, which cannot be explained
 by their statistical uncertainties. In our opinion, such discrepancies
 could be caused by several factors. There was a significant difference
 in the trigger of the two experiments and in the energy resolution
 of the calorimeters, as the experiment with the EPT (Run-II) was made
 with a much higher MAMI electron current. A poorer energy resolution in
 Run-II resulted in a larger smearing of the $\cos\theta$ angular dependence. 
 Another possible explanation is a slight mismatch in the energy calibrations
 of the two different tagged-photon spectrometers.
 Overall, the average systematic uncertainty
 in the differential cross sections due to the resolution in
 the c.m. $\theta$ and possible imperfections in the angular acceptance
 was estimated to be 2\% for the results of Run-I. The corresponding systematic
 uncertainty for the results of Run-II is at least twice as high.
\begin{figure*}
\includegraphics[width=0.97\textwidth,height=6.7cm]{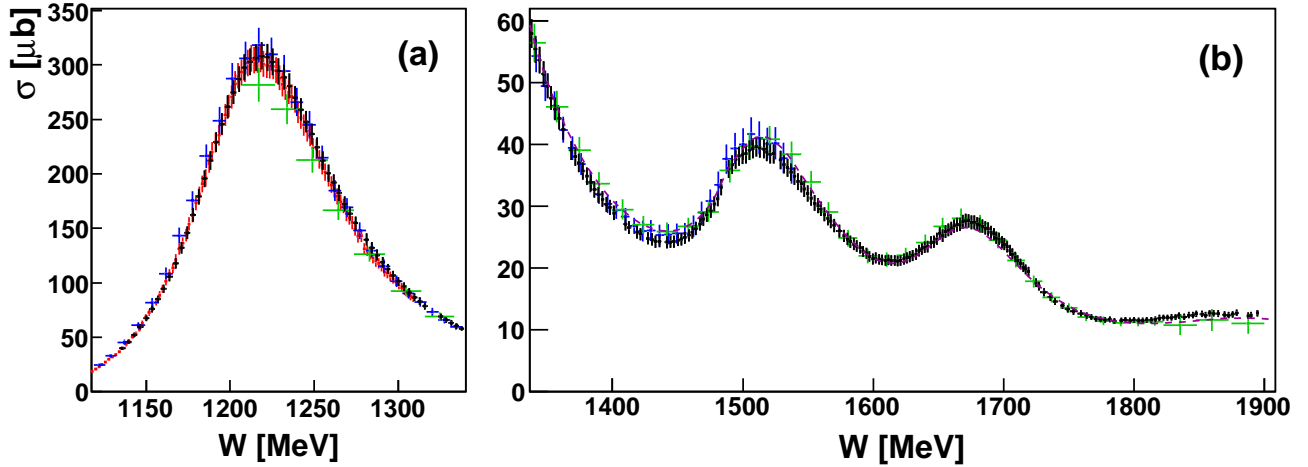}
\caption{(Color online)
 Total cross sections for $\gamma p\to \pi^0 p$ from this work
 (black points) as a function of the c.m. energy $W$ compared with
 earlier A2 results~\protect\cite{hornidge13} (red points),
 with results from Refs.~\protect\cite{be06} (blue points)
 and ~\protect\cite{ba05} (green points), and with
 the SAID solution CM12~\protect\cite{cm12} shown in ~(b)
 by a magenta dashed line. For a better comparison, the energy range
 is divided in two intervals: (a) $W<1340$~MeV and (b) $W>1340$~MeV.
 The error bars on the experimental data points represent their
 total uncertainties.
}
 \label{fig:tcs_pi0p_a2}
\end{figure*}

 Comparison of the results for the $\gamma p\to \pi^0 p$ differential
 cross sections from the present work with the most recent published data
 from other collaborations~\cite{be06,Bartalini_05,du07,ba05,cr11}
 is illustrated in Fig.~\ref{fig:dcs_pi0p_a2_others}
 for eight different energies. Previous results with either large
 angular bins or very restricted angular coverage are not included
 in the comparison. As seen in Fig.~\ref{fig:dcs_pi0p_a2_others},
 data with full angular coverage existed only in the region
 below the third resonance region. The results of this work
 are in a general agreement with all the data shown, although those
 data are not always entirely consistent with each other.  
 As also evident, the present data from Run-I have much superior
 statistical accuracy, combined with a finer angular and energy binning.
\begin{figure*}
\includegraphics[width=0.96\textwidth,height=19.cm]{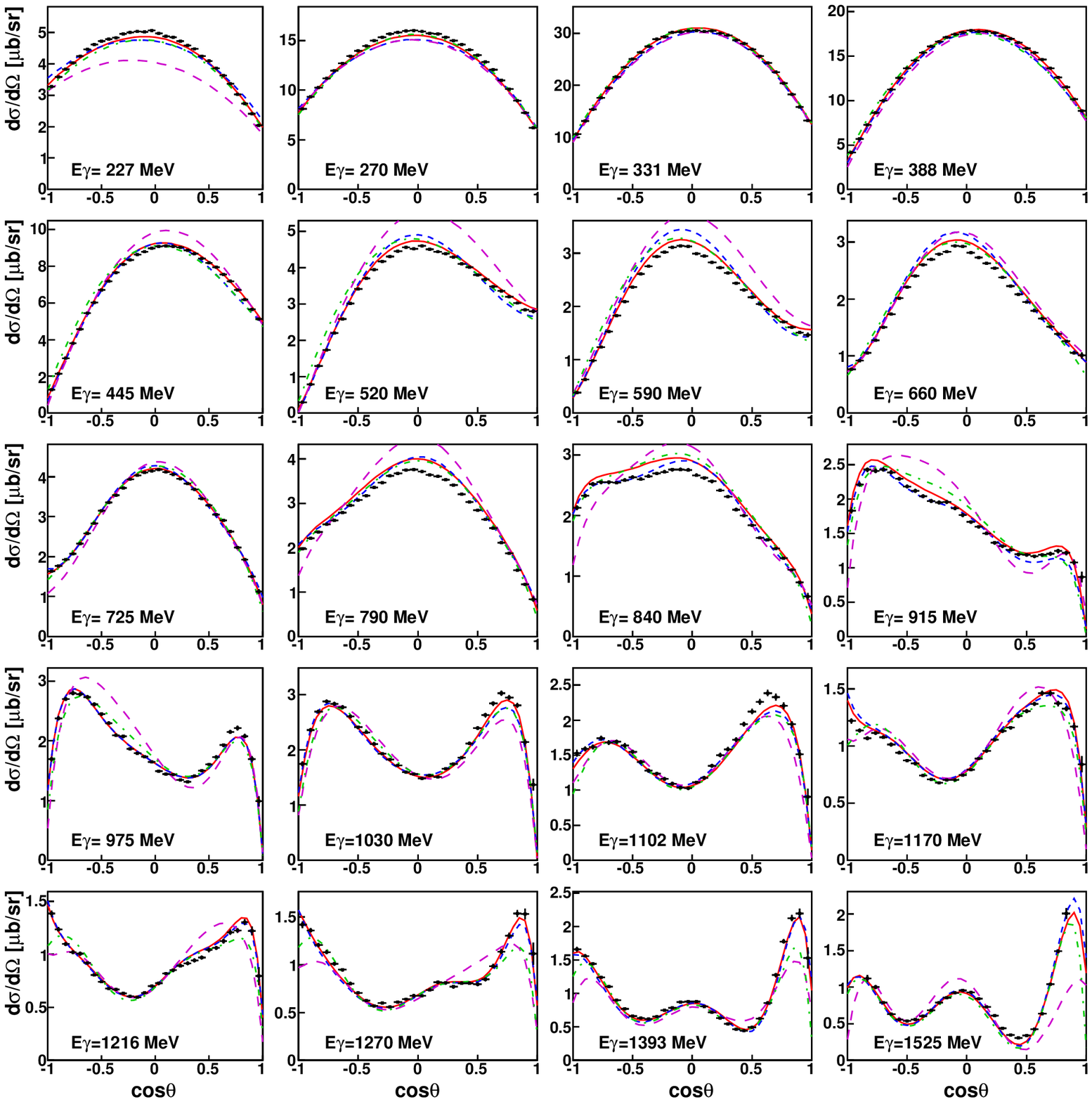}
\caption {(Color online)
 Selected results of this work (black points)
 for the $\gamma p\to \pi^0 p$ differential cross sections
 compared to existing PWA solutions from
 SAID CM12~\protect\cite{cm12} (blue dashed line),
 MAID2007~\protect\cite{maid} (magenta long-dashed line),
 and Bonn-Gatchina BG2014-02~\protect\cite{bnga} (green dash-dotted line) and to
 a new SAID PR15 solution (red solid line) obtained after adding
 the present data points into the fit. The error bars on all data points
 represent statistical uncertainties only.
}
\label{fig:dcs1_pi0p_a2_pwa}
\end{figure*}

The total cross sections for $\gamma p\to \pi^0 p$ were determined
by integrating the differential cross sections for the data obtained
from Run-I, which have the full angular coverage.
 The resulting total cross sections are plotted in
 Fig.~\ref{fig:tcs_pi0p_a2} as a function of the c.m. energy $W$
 along with earlier A2 results obtained for energies up to
 $E_\gamma=444$~MeV ($W=1309$~MeV)~\cite{hornidge13} and results from
 Ref.~\protect\cite{be06}, obtained for energies up to
 $E_\gamma=790$~MeV ($W=1537$~MeV), and from Ref.~\cite{ba05},
 obtained for energies above $E_\gamma=300$~MeV ($W=1201$~MeV).
 The latter data have no full angular coverage; they were obtained
 with extrapolating experimental differential cross sections into
 forward and backward angles by using the corresponding PWA results.
 For completeness, the data at higher energies are also compared
 in Fig.~\ref{fig:tcs_pi0p_a2}(b) to the SAID solution
 CM12~\cite{cm12}, shown by a magenta dashed line.
 As seen in Fig.~\ref{fig:tcs_pi0p_a2}(a) and (b),
 the present results are in good agreement, within
 the total uncertainties (calculated by adding
 the statistical and the overall systematic uncertainty
 in quadrature), with the earlier A2 results
 and in general agreement with the most recent
 experimental data and SAID CM2 solution based on
 fitting previous data, even if they do not have
 the full angular coverage.
\begin{table*}
\caption
[tab:pwachi2]{
 Average $\chi^2/\mathrm{dp}$ values (including information
 on the total $\chi^2$ from Eq.~(\protect\ref{eq:chi2}) and
 the number of data points used)
 for MAID2007~\protect\cite{maid} (fitting data only below $W=2$~GeV)
 and three SAID solutions: CM12~\protect\cite{cm12},
 DU13~\protect\cite{du13}, and PR15.
} \label{tab:chi2}
\begin{ruledtabular}
\begin{tabular}{|c|c|c|c|c|}
\hline
 $\chi^2/\mathrm{dp}$
 & MAID2007 & SAID CM12 &  SAID DU13 & SAID PR15 \\
\hline
 present $\gamma p\to\pi^0p$
 & $99722/7978=12.5$ & $24036/7978=3.0$ & $28745/7978=3.6$  & $9434/7978=1.2$ \\
\hline
 previous & & & & \\
 $\gamma p\to\pi^0p$
 & $135195/15468=8.7$ & $73211/17087=4.3$ & $41500/17087=2.4$  & $40793/17087=2.4$ \\
 $\gamma p\to\pi^+n$
 & $171853/8092=21.2$ & $23533/8959=2.6$ & $19312/8959=2.2$  & $17540/8959=2.0$ \\
 $\gamma n\to\pi^-p$
 & $25335/2806=9.0$ & $53657/3162=17.0$ & $7561/3162=2.4$  & $6572/3162=2.1$ \\
 $\gamma n\to\pi^0n$
 & $22568/364=62.0$ & $979/364=2.7$ & $1436/364=4.0$  & $1187/364=3.3$ \\
\hline
\end{tabular}
\end{ruledtabular}
\end{table*}
\begin{figure*}
%\centering
\begin{minipage}[c]{.31\textwidth}
\centering
\includegraphics[height=5.2cm,width=5.0cm,angle=90]{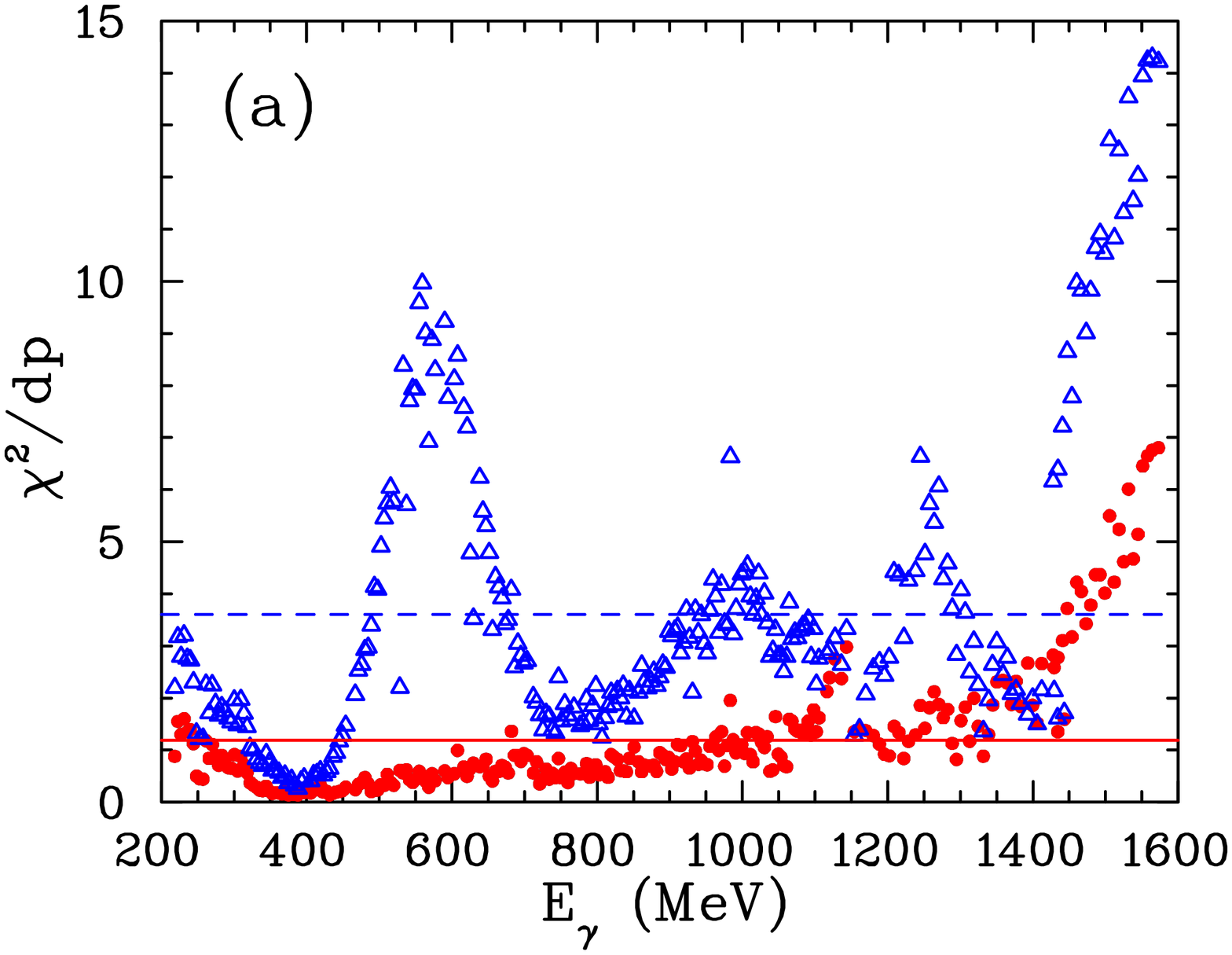}
\end{minipage}
\begin{minipage}[c]{.31\textwidth}
\includegraphics[height=5.2cm,width=5.0cm,angle=90]{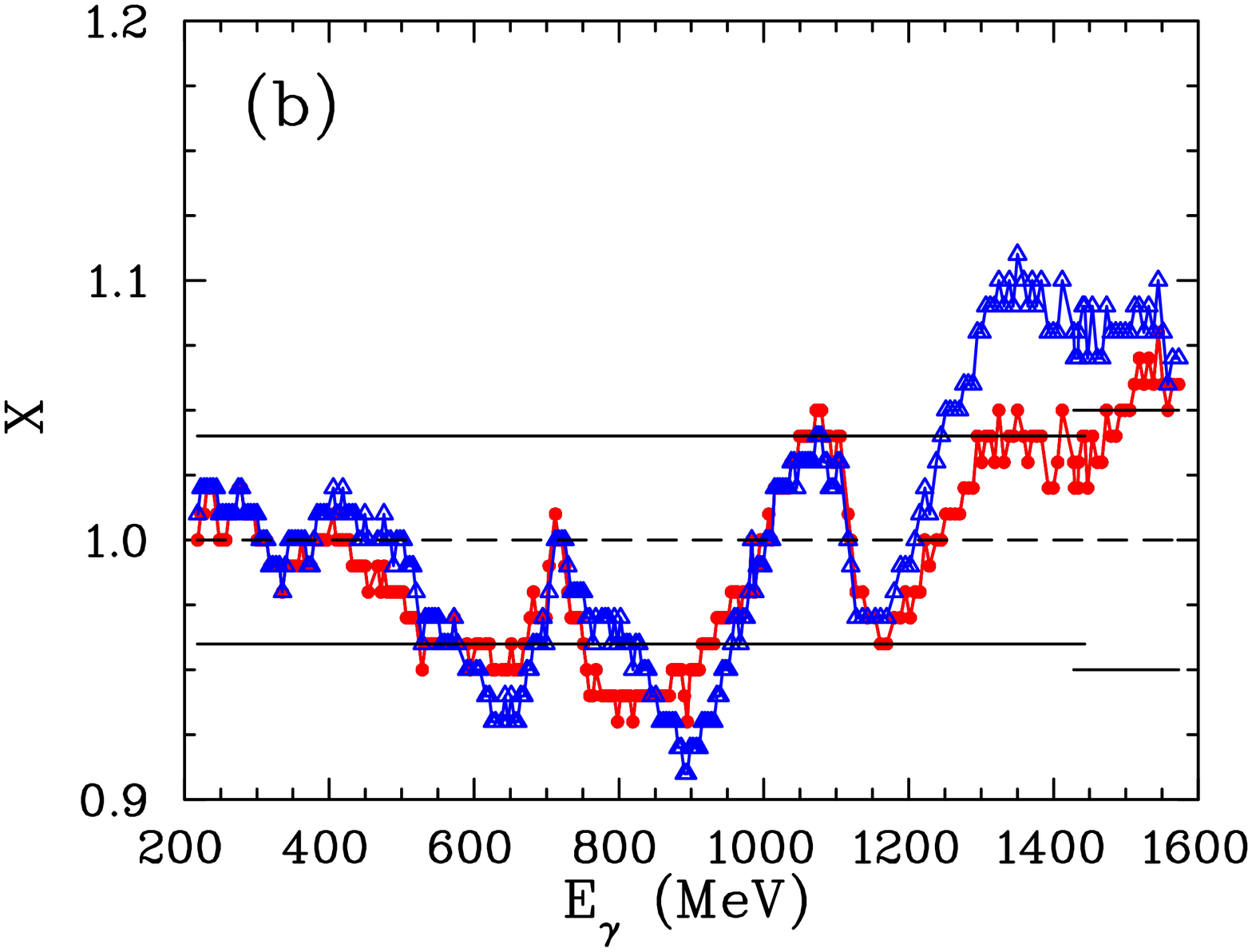}
\end{minipage}
\begin{minipage}[c]{.34\textwidth}
\centering
\caption{(Color online)
  Comparison of the previous SAID solution DU13~\protect\cite{du13}
 (blue open triangles) applied to the present data and the new SAID PR15 solution
 (red full circles) obtained after adding the present data into the fit for:
 ~(a)~$\chi^2/\mathrm{dp}$ values averaged only within each energy bin
      $E_\gamma$, where the horizontal lines (blue dashed for DU13 and
     red solid for PR15) show the corresponding $\chi^2/\mathrm{dp}$ values
     from Table~\protect\ref{tab:chi2};
 ~(b)~results for the normalization parameter $X$ as a function
      of $E_\gamma$, where the horizontal solid lines show the deviations
     (from the dashed line at $X=1$) equal to the magnitude of the overall
     systematic uncertainties, which are 4\% and 5\% for the data
     from Run-I and Run-II, respectively.
}
 \label{fig:chi2_vs_eg}
\end{minipage}
\end{figure*}

 Comparison of the present results for the $\gamma p\to \pi^0 p$ differential
 cross sections with predictions of different partial-wave and
 coupled-channel analyses is illustrated for selected energies
 in Fig.~\ref{fig:dcs1_pi0p_a2_pwa}. Three sets of predictions shown
 in this figure correspond to PWAs (namely, SAID CM12~\cite{cm12},
 MAID2007~\cite{maid}, and Bonn-Gatchina BG2014-02~\cite{bnga})
 based on fitting previous data. The fourth set of curves comes from
 SAID PR15 solution obtained after adding the present data points
 into the SAID fit. Compared to the SAID solution CM12, the new fit PR15
 also involves the data added to obtain SAID DU13 solution~\cite{du13}
 and other most recent polarization data from
 CBELSA/TAPS~\cite{Hartmann_14,Gottschall_14} and A2 at MAMI~\cite{Sikora_14}.
 As seen, all three the previous PWAs are very close to
 each other and to the present data points in the energy
 region around the $\Delta(1232)3/2^+$ maximum. Differences between
 the PWAs themselves as well as between the PWAs and the data points
 increase, especially for the MAID2007 solution, at energies
 that are away from the $\Delta(1232)3/2^+$ peak. Then all the PWAs
 and the data points become closer again in 
 the second and the third resonance regions.
 The largest difference between the present data and the three
 previous PWAs is observed for the MAID2007 solution; this could
 be explained by the fact that many of the most recent data sets were
 not available at the time when the MAID2007 solution was released.

 Quantitatively, the agreement of the present data with PWAs
 can usually be estimated via $\chi^2$ values calculated
 from deviations of experimental data points from PWA solutions.
 The $\chi^2$ function used in the minimization
 procedure by SAID is given by formula~\cite{arndt02}
\begin{equation}
	\chi^2 = \sum_i ( \frac{X \Theta_i - \Theta^{\rm exp}_i}{\epsilon_i} )^2
	+ ( \frac{X-1}{\epsilon_X} )^2,
\label{eq:chi2}
\end{equation}
 where $\Theta^{\rm exp}_i$ is an individual experimental value
 of the differential cross section with its uncertainty
 $\epsilon_i$ (includes both the statistical and the individual
 systematic uncertainty discussed in the text above),
 $\Theta_i$ is the fit value calculated for the same energy
 and the production angle, and $X$ is a normalization
 parameter with its uncertainty $\epsilon_X$ that
 is determined by the overall systematic uncertainties of the data
 (see start of this section).
 The average $\chi^2/\mathrm{dp}$ values, along with the information
 on the total $\chi^2$ from Eq.~(\ref{eq:chi2}) and
 the number of data points used (dp), are listed
 in Table~\ref{tab:chi2} for MAID2007~\cite{maid} and three SAID
 solutions: CM12~\cite{cm12}, DU13~\cite{du13}, and PR15.
 As seen, the numbers in Table~\ref{tab:chi2} confirm that
 the average $\chi^2/\mathrm{dp}$ is worse for
 the older solution MAID2007 than for the more recent SAID solutions
  CM12 and DU13. The magnitudes of $\chi^2/\mathrm{dp}$ obtained
 for the new SAID solution PR15 indicate that the new $\gamma p\to \pi^0 p$
 data are consistent with the existing $\gamma N\to\pi N$ data sets.
 The number of data points available for each photoproduction
 reaction, which are listed in Table~\ref{tab:chi2}, shows that
 the present $\gamma p\to\pi^0p$ data increase the existing
 $\gamma p\to\pi^0p$ statistics by $\sim 47\%$ and
 the existing $\gamma N\to\pi N$ statistics by $\sim 27\%$.

 The $\chi^2/\mathrm{dp}$ values, averaged within each
 energy bin, are plotted as a function of $E_\gamma$
 in Fig.~\ref{fig:chi2_vs_eg}(a) for the previous SAID
 solution DU13~\cite{du13} and for the new SAID PR15 solution,
 obtained after adding the present data into the fit.
 This comparison demonstrates the energies at which the new data
 give the most significant impact.
 The results obtained for the normalization parameter $X$
 from the DU13 and PR15 solutions are plotted 
 as a function of $E_\gamma$ in Fig.~\ref{fig:chi2_vs_eg}(b).
\begin{figure*}
\includegraphics[width=0.95\textwidth,height=19.cm]{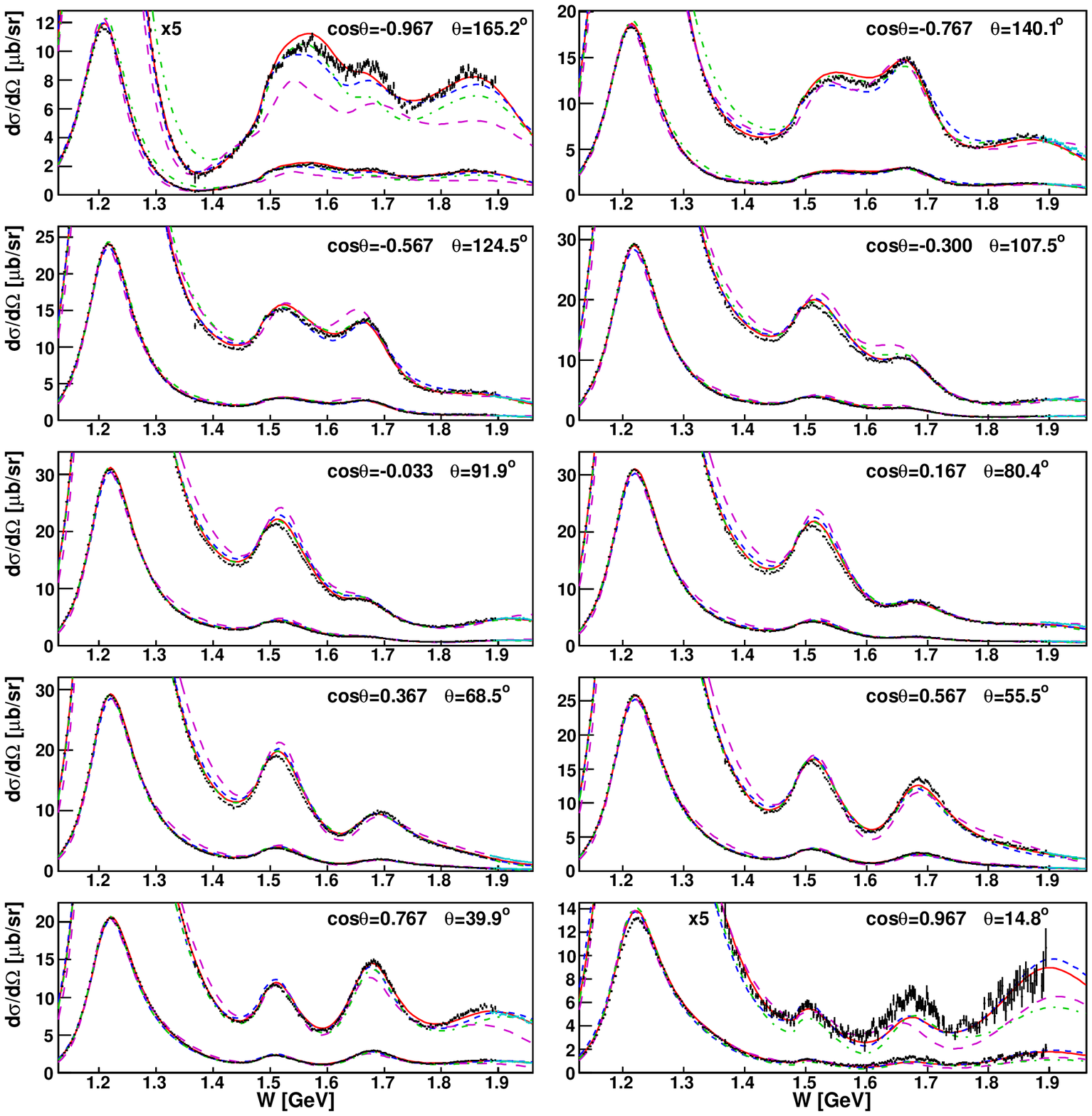}
\caption {(Color online)
 Present results for the $\gamma p\to\pi^0p$ excitation function
 as a function of the c.m. energy $W$, shown
 for ten of 30 measured $\cos\theta$ bins (where
 $\theta$ is the $\pi^0$ production angle),
 are compared to previous PWA solutions from
 SAID CM12~\protect\cite{cm12} (blue dashed line),
 MAID2007~\protect\cite{maid} (magenta long-dashed line),
 and BG2014-02~\protect\cite{bnga} (green dash-dotted line) and to
 a new SAID PR15 solution (red solid line). 
 The error bars on all data points (black points for Run-I and
 cyan points for Run-II) represent statistical uncertainties only.
 For a better comparison, all distributions are also plotted
 after rescaling by a factor of 5.
}
\label{fig:pi0p_dcs_theta}
\end{figure*}

 Another useful presentation of the new $\gamma p\to\pi^0p$ results is
 obtained by plotting them as an excitation function at a particular
 production angle $\theta$ of $\pi^0$. In Fig.~\ref{fig:pi0p_dcs_theta},
 the differential cross sections are shown as a function of
 the c.m. energy $W$ for ten of 30 measured $\cos\theta$ bins.
 As seen, the peaks corresponding to the first, the second, and
 the third resonance region are clearly evident in the excitation
 function for each $\pi^0$ production angle.
 The experimental distributions are also compared to the three
 previous PWA solutions (SAID CM12~\cite{cm12}, MAID2007~\cite{maid},
 and BG2014-02~\cite{bnga}) and the new SAID PR15 solution.
 For a better comparison in the energy range away from the $\Delta(1232)3/2^+$
 peak, all distributions are also plotted after rescaling by a factor of 5.
 The comparison demonstrates that the largest discrepancies are
 seen for the very forward and backward production angles.
\begin{figure*}
%\centering
\begin{minipage}[c]{.32\textwidth}
\centering
\includegraphics[height=5.3cm,width=5.0cm,angle=90]{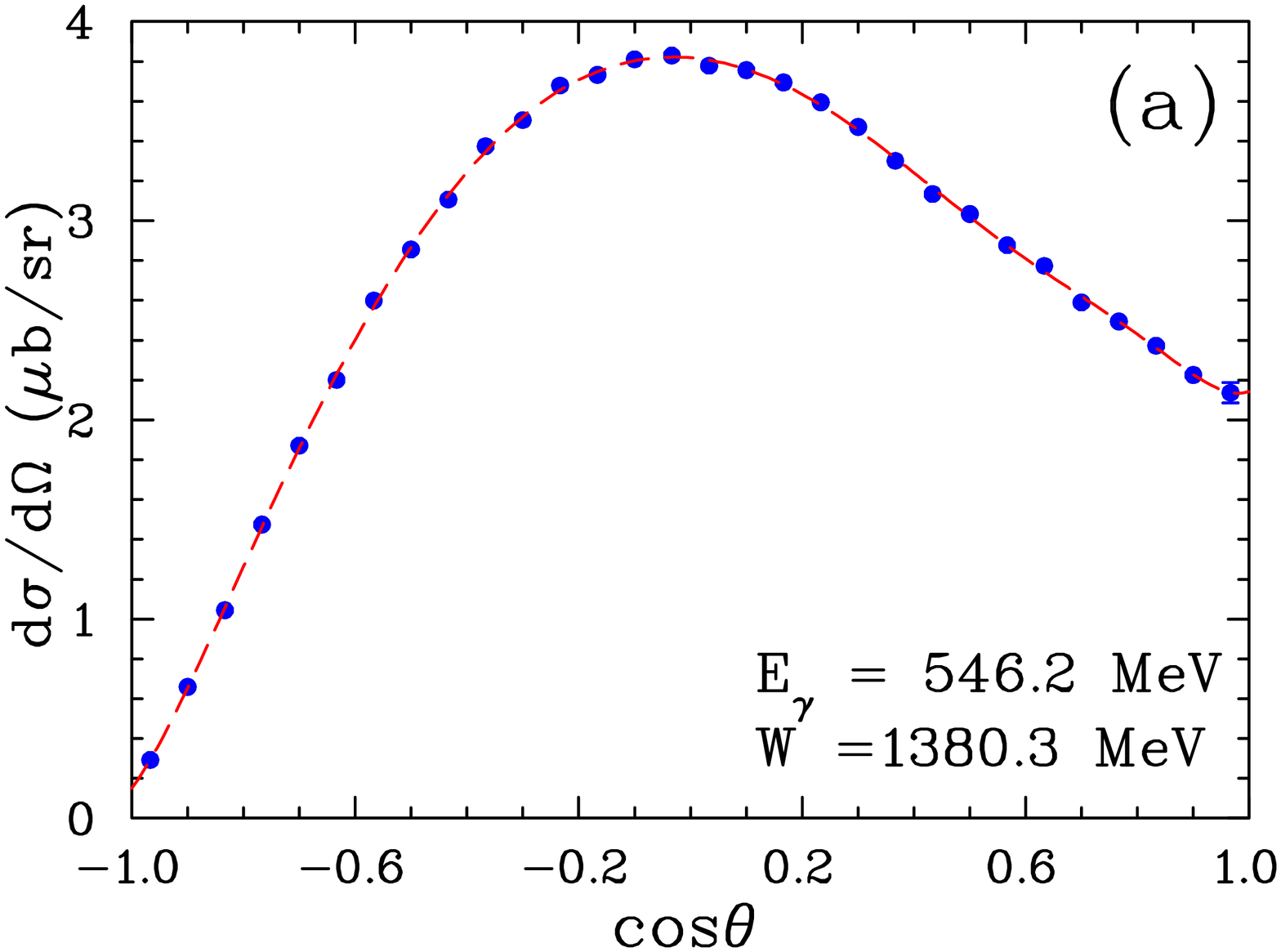}
\end{minipage}
\begin{minipage}[c]{.32\textwidth}
\centering
\includegraphics[height=5.3cm,width=5.0cm,angle=90]{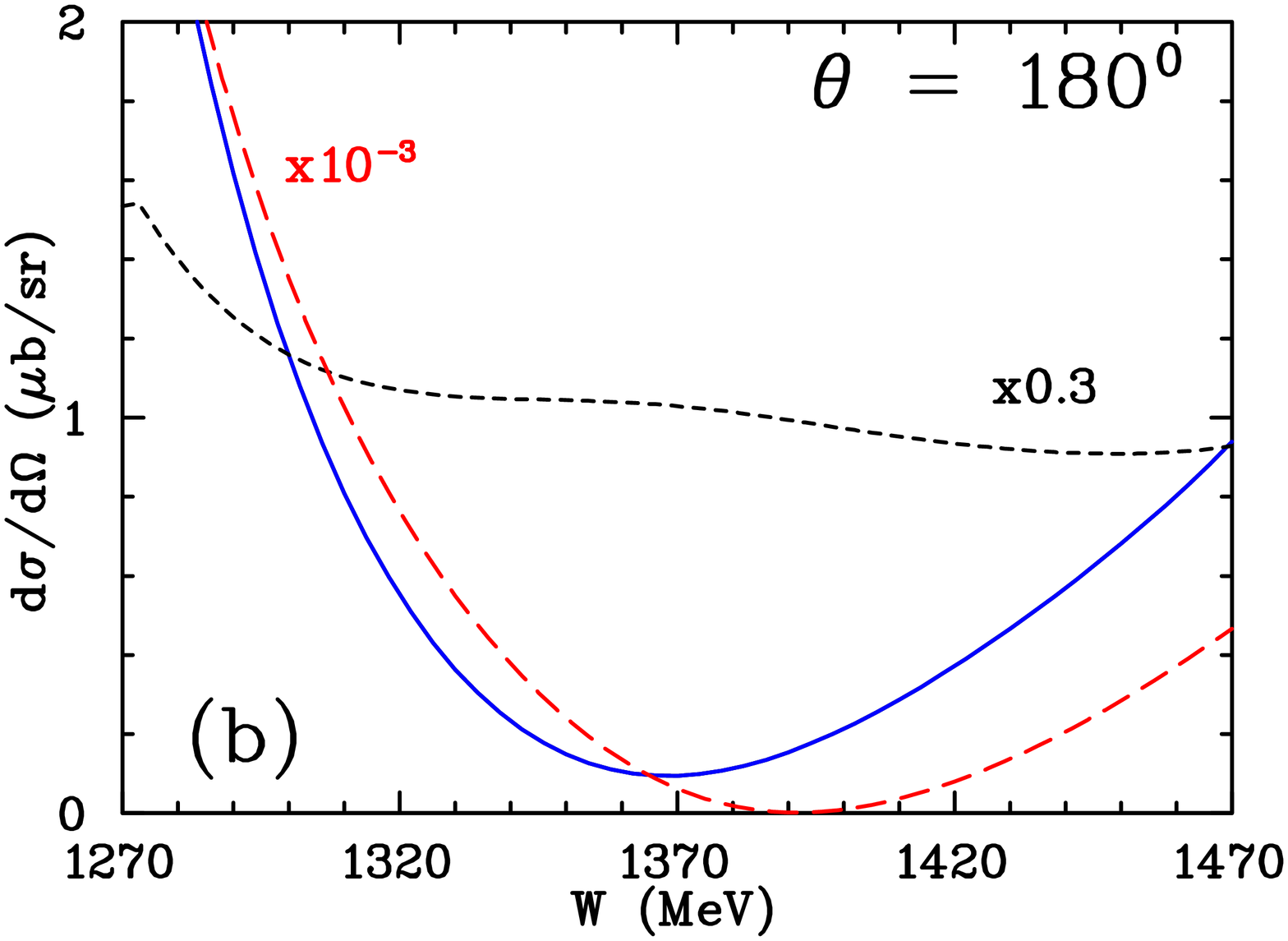}
\end{minipage}
\begin{minipage}[c]{.32\textwidth}
\centering
\caption{(Color online)
 ~(a)~ present results for the $\gamma p\to \pi^0 p$ differential cross section
  (blue full circles) at $E_\gamma=546$~MeV ($W=1380$~MeV) fitted
 with Legendre polynomials (red dashed line);
 ~(b)~ the new SAID PR15 solution (blue solid line) for
  $\pi^0$ production angle $\theta=180^{\circ}$ ($\cos\theta=-1$)
  as a function of the c.m. energy $W$, compared with
 a similar $\pi^- p\to \pi^0 n$ prediction (rescaled by a factor $10^{-3}$)
 from SAID WI08~\protect\cite{SAID_WI08}
 (red dashed line) and a $\gamma p\to\pi^+n$ prediction
 (rescaled by a factor 0.3) from SAID PR15 (black dashed line).
}
 \label{fig:dcs_pi0p_th180}
\end{minipage}
\end{figure*}

 The strongest dip in the $\gamma p\to\pi^0p$ excitation
 function is observed at very backward $\pi^0$ angles at the c.m.\ energies
 slightly below $W=1.4$~GeV. An example of the $\gamma p\to\pi^0p$
 differential cross section near this dip (namely, at $W=1380$~MeV)
 is shown in Fig.~\ref{fig:dcs_pi0p_th180}(a).
 This effect, caused by the nearly complete cancellation
 between different contributions to the production cross section,
 is even better seen in the results of the new SAID PR15 solution,
 shown in Fig.~\ref{fig:dcs_pi0p_th180}(b) by a blue solid line
 for $\cos\theta=-1$ (or $\theta=180^{\circ}$).
 It is interesting that a very similar cancellation
 in the very backward $\pi^0$ production angles
 was observed in the charge-exchange reaction $\pi^- p\to \pi^0 n$ at
 the same energy range. The SAID WI08 solution~\cite{SAID_WI08}
 (rescaled by a factor $10^{-3}$) for $\pi^- p\to \pi^0 n$ at
 $\theta=180^{\circ}$ is plotted in Fig.~\ref{fig:dcs_pi0p_th180}(b)
 by a red dashed line.
 In contrast, the $\gamma p\to\pi^+n$ reaction, depicted for
 the SAID PR15 solution by a black dashed line,
 has no such feature. Unfortunately, the quality and the angular
 coverage of the existing $\gamma n\to\pi^0n$ data do not allow
 a reliable prediction for very backward production $\pi^0$ angles
 in this energy range~\cite{Dieterle_14}.

 For the c.m.\ energies above $W=1.4$~GeV, the present results
 for very backward $\pi^0$ angles are in good agreement with
 similar results ($\theta=170^{\circ}$) published in Ref.~\cite{Althoff_79},
 also showing a cusp structure at the $\eta$-photoproduction threshold
 ($W=1.486$~GeV).  

 The statistical accuracy of the present results is sufficient
 for reliable fits of the $\gamma p\to\pi^0p$ differential cross sections
 with series of Legendre polynomials up to a high order.
 Of course, such decompositions are much less reliable without full angular
 coverage of the $\pi^0$ production angle. Thus, the present results
 from Run-II should be used only in energy-dependent
 PWAs with additional model assumptions.
 For Run-I, we describe the $\gamma p\to\pi^0p$ differential cross sections
 of each energy bin in terms of a series of orthogonal Legendre
 polynomials $P_j(\cos\theta)$:
\begin{equation}
        \frac{d\sigma}{d\Omega}(W, \cos\theta) = \sum\limits_{j=0}^{j_{\rm max}}~
        A_j(W)~P_j(\cos\theta) ,
\nonumber
\end{equation}
with the energy-dependent decomposition coefficients $A_j(W)$.

 From the phenomenological point of view, the number
 of terms required in the series, $j_{\rm max}$, is related
 to the angular momenta of essential partial-wave amplitudes
 or to the highest spin of resonances, existing in
 the studied energy range.
 An isolated resonance with spin $J = (2\ell+1)/2$ can contribute
 to coefficients $A_j$ only with even $j$ up to $j = 2\ell$
 and, hypothetically, to other coefficients via interference with
 sizable background amplitudes.
 For example, a wide four-star resonance $\Delta(2420)$ 
 with $J^P=11/2^+$~\cite{PDG}, although lying  above the energy range
 of the present measurements, can affect $A_j$ up to $j=10$ by itself
 and even higher $A_j$ via the interference with other states.
 On the other hand, the maximum order $j_{\rm max}$, up to which
 such a decomposition is meaningful, is limited by the data quality.
\begin{figure}[hbt]
\includegraphics[width=.45\textwidth]{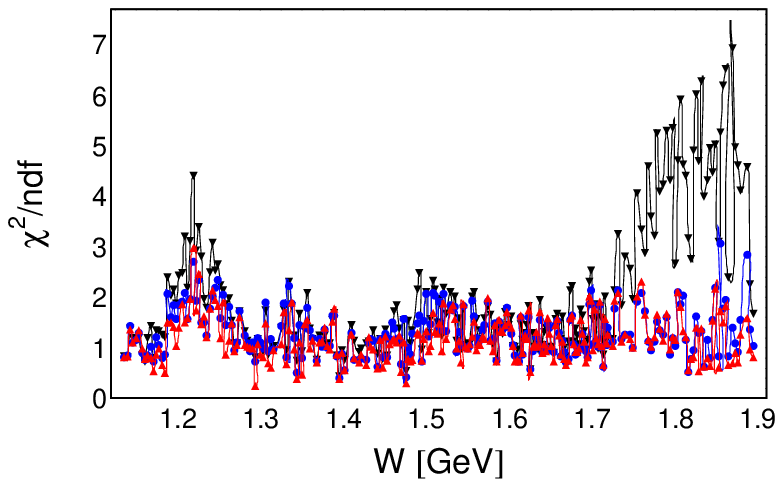}
\caption {(Color online)
 Energy dependence of the reduced $\chi^2$ values for Legendre fits
 of the present $\gamma p\to\pi^0p$ differential cross sections
 with $j_{\rm max}=6$ (black triangles down), 8 (blue circles), and 10 (red triangles up),
 where the angular-dependent systematic uncertainties were not
 taken into account.
}
\label{fig:leg_fit_chi2}
\end{figure}
\begin{figure*}
\includegraphics[width=.95\textwidth]{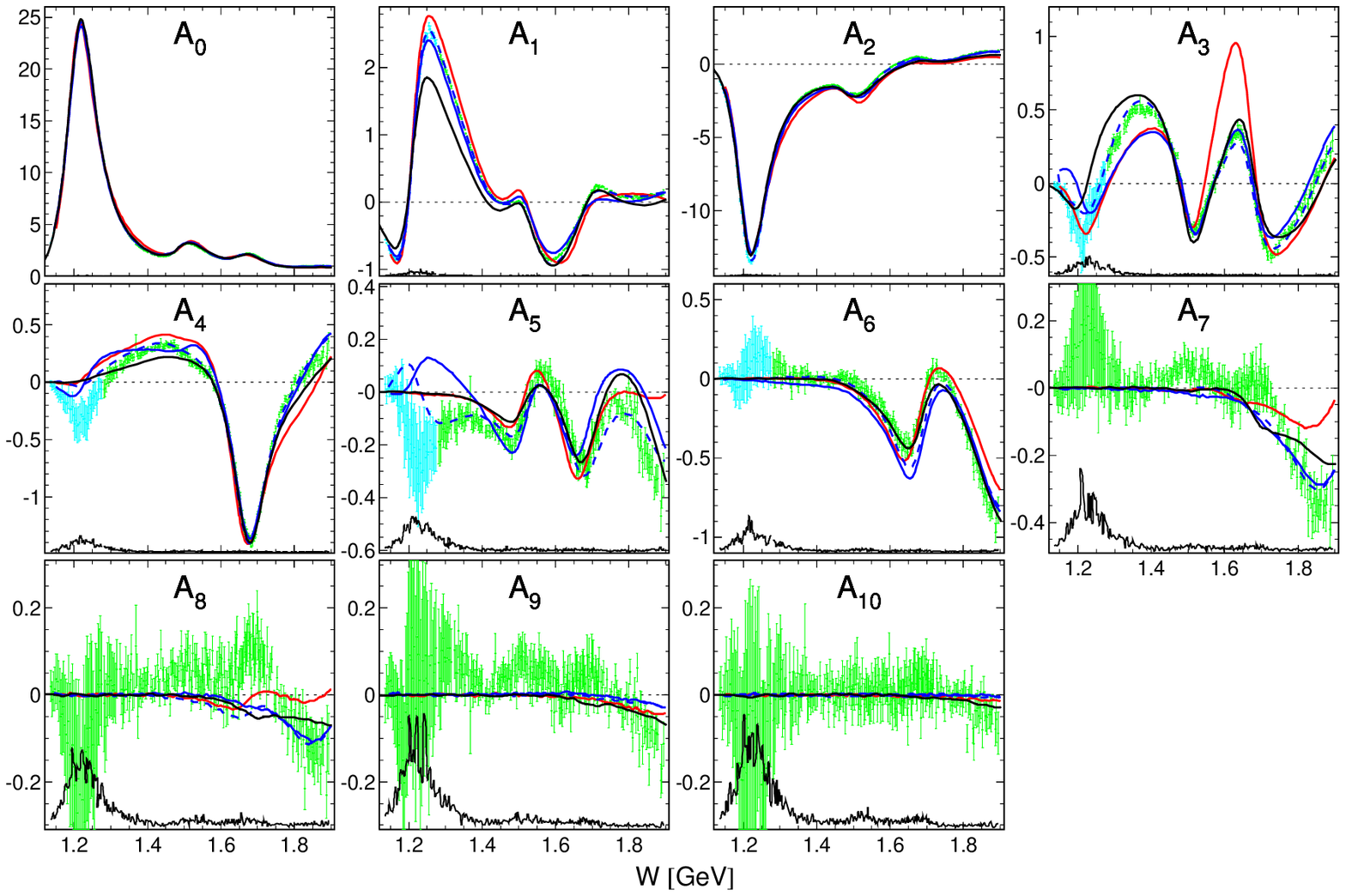}
\caption {(Color online)
 Coefficients $A_j$ (in $\mu\mathrm{b/sr}$ units) obtained from
 fitting the present $\gamma p\to\pi^0p$
 differential cross sections with Legendre polynomials
 up to order six (cyan points) and ten (green points),
 compared to the results of the Legendre-polynomial fits
 ($j_{\rm max} = 10$) made to the predictions generated from
 SAID CM12~\protect\cite{cm12} (blue solid line),
 MAID2007~\protect\cite{maid} (red solid line),
 BG2014-02~\protect\cite{bnga} (black solid line), and
 SAID PR15 (blue dashed line).
 The error bars on all points represent $A_j$ uncertainties
 from the fits with using only the statistical uncertainties.
 A black histogram at the bottom of each plot shows the
 systematic uncertainty in $A_j$ caused by the angular-dependent
 systematics in the differential cross sections.
}
\label{fig:dcs_pi0p_leg_coef}
\end{figure*}

In this work, the data of Run-I were fitted up to order $j_{\rm max} = 10$.
It was observed that including Legendre polynomials beyond $j_{\rm max} = 8$
in the fits could not significantly improve the reduced $\chi^2$
(i.e., divided by the number of degrees of freedom, $\chi^2/{\rm ndf}$).
The comparison of the reduced $\chi^2$ for the fits of the data
from Run-I with $j_{\rm max} = 6$, 8, and 10 is shown in
Fig.~\ref{fig:leg_fit_chi2}, using only the statistical uncertainties.
In the energy range around the $\Delta(1232)3/2^+$ region, which is below $W=1.3$~GeV,
fitting with $j_{\rm max}=6$ should, in principle, be sufficient for a fairly good
description of the differential cross sections. However,
the high statistical accuracy, which arises from the very large
cross sections, results in the reduced $\chi^2$ values varying mostly
between two and three. As also seen in Fig.~\ref{fig:leg_fit_chi2},
 including higher-order coefficients in the fits does not improve
 the situation much. It is necessary to take into account angular-dependent
 systematic uncertainties, with values varying between 1\% and 2\%,
 in order to obtain reasonable $\chi^2$ values in the $\Delta(1232)3/2^+$
 region, even in the fits with $j_{\rm max}=6$.
 Above $W=1.7$~GeV, using $j_{\rm max}=6$ is not sufficient, and
 the data require the minimum order of 
 $j_{\rm max} = 8$, corresponding to $G$-wave amplitudes.

The results of the Legendre-polynomial fits for each coefficient $A_j$
are depicted in Fig.~\ref{fig:dcs_pi0p_leg_coef}, showing their
energy dependence in unprecedented resolution.
The $A_j$ error bars represent uncertainties obtained
from fits that took into account only the statistical uncertainties
in the cross sections.
As previously discussed, the effect from the angular-dependent systematic
uncertainties should be considered, especially for the $\Delta(1232)3/2^+$
energy region. To estimate the impact from these uncertainties, the values
of the differential cross section for each angular bin were randomly shifted
within the uncertainty magnitude (2\%), followed by a new
Legendre-polynomial fit. Such a procedure was then repeated several times,
and the average spread of the fit results for each $A_j$ was considered
as its systematic uncertainty caused by the angular-dependent systematic
uncertainties in the differential cross sections.
The systematic uncertainties in $A_j$, obtained from the fits with
 $j_{\rm max} = 10$, are depicted by a black histogram at the bottom of each
 plot in Fig.~\ref{fig:dcs_pi0p_leg_coef}, clearly peaking at the position of
$\Delta(1232)3/2^+$. In this region, the magnitudes of the systematic uncertainties
 in $A_j$ at high orders are very close to the magnitudes of $A_j$ themselves.
 The Legendre-polynomial fits for the c.m. energies
below $W=1280$~MeV were made for both $j_{\rm max}=6$ and $j_{\rm max} = 10$.
Their comparison showed good agreement of the results for $A_{0}$ to $A_{6}$.

 One of the most obvious features in the energy
 dependence of $A_j$ is the absence of any well-defined sharp structures,
 which could be associated with unknown narrow resonances.
 The results for $A_{0}$ show good agreement
 with the values for the total cross sections obtained by the
 integration of the differential cross sections, providing
 another confirmation of the good quality of the present data.
 The $A_{0}$ energy dependence clearly demonstrates
 the first, second, and third resonance regions, but appears
 structureless at higher energies. In contrast, some of the
 other $A_j$ reveal structures even at the energies above
 the third resonance region.
 These structures could be associated with contributions from $N^*$ and/or
 $\Delta^*$ states, with spins up to 7/2 or higher, that are known in this 
 region or near it~\cite{PDG}. These resonances have masses close to
 the upper end of the studied energy range, as $\Delta(1950)7/2^+$,
 or even above it, as $N(2190)7/2^-$, $N(2220)9/2^+$, and
 $N(2250)9/2^-$, but their large widths ($\sim 300$~MeV or even larger)
 make those resonance contributions noticeable in the present data.

 In Fig.~\ref{fig:dcs_pi0p_leg_coef}, the coefficients $A_j$
 are also compared to results obtained from
 the predictions generated with SAID CM12~\cite{cm12},
 MAID2007~\cite{maid}, BG2014-02~\cite{bnga}, and SAID PR15,
 by fitting them, after adding small uncertainties,
 with Legendre polynomials up to $j_{\rm max} = 10$.
 As seen, the Legendre-polynomial fits to the predictions of
 all models reproduce fairly well only the coefficients
 $A_0$ and $A_2$. Partial agreement is observed for
 $A_1$ and the coefficients $A_3$ to $A_7$.
 The coefficients $A_8$ to $A_{10}$ are small and show some
 meaningful structures only above $W=1.7$~GeV.

 Of course, such a Legendre decomposition of the present
 differential cross sections is only a very first step towards
 a full single-energy partial-wave analysis.
 A direct extraction of partial-wave amplitudes by simultaneous
 fitting the present differential cross sections and polarization observables
 will be performed in the future. However, the present analysis already shows
 the sensitivity of the new data to partial-wave amplitudes
 up to $G$ waves ($\ell = 4$). 

\section{Summary and conclusions}
\label{sec:Conclusion}

 New measurements of $\pi^0$ photoproduction on the proton have
 been conducted with the A2 tagged-photon facilities at energies
 provided by the Mainz Microtron, MAMI C, with a maximum
 electron-beam energy of 1.6~GeV.
 The differential cross sections, obtained with a fine energy
 and angular binning, increase the existing quantity of
 $\pi^0$ photoproduction data by $\sim 47\%$.
 A new solution PR15 of the SAID partial-wave analysis obtained
 after adding the new data into the fit is presented.
 The magnitudes of $\chi^2/\mathrm{dp}$ obtained
 for this solution indicate that the new $\gamma p\to \pi^0 p$
 data are consistent with the existing $\gamma N\to\pi N$ data sets.
 Owing to the unprecedented statistical accuracy 
 and the full angular coverage, the results are sensitive to high
 partial-wave amplitudes. This is demonstrated by the decomposition
 of the differential cross sections in terms of Legendre polynomials
 and by further comparison to model predictions.
 The present data are expected to be invaluable for future partial-wave 
 and coupled-channel analyses, which  could provide much stronger 
 constraints on the properties of nucleon states known in this energy 
 range and, perhaps, even reveal new resonances.
 A more detailed analysis of the present $\gamma p\to\pi^0 p$
 differential cross sections, combined with results from
 the A2 Collaboration obtained by measuring 
 polarization observables, is currently in progress and
 will be published separately.

%\section*{Acknowledgments}
\begin{acknowledgments}
We thank A.~Donnachie for useful remarks and continuous interest in the paper.
The authors wish to acknowledge the excellent
support of the accelerator group and operators of MAMI. This work was
supported by the Deutsche Forschungsgemeinschaft (SFB443, SFB/TR16, and
SFB1044), the European Community-Research
Infrastructure Activity under the FP6 ``Structuring the European Research Area"
programme (Hadron Physics, Contract No. RII3-CT-2004-506078),
Schweizerischer Nationalfonds (Contract Nos. 200020-156983,
132799, 121781, 117601, 113511), the UK Science and
Technology Facilities Council (STFC 57071/1, 50727/1),
the U.S. Department of Energy (Offices of Science and Nuclear Physics,
 Award Nos. DE-FG02-99-ER41110, DE-FG02-88ER40415, DE-FG02-01-ER41194)
 and National Science Foundation (Grant Nos. PHY-1039130, IIA-1358175),
 INFN (Italy), and NSERC (Canada). Ya.~I.~A. acknowledges support by
the Russian Science Foundation (Grant No. 14-22-00281).
A.~F. acknowledges additional support from the TPU (Grant No. LRU-FTI-123-2014)
 and the MSE Program ``Nauka'' (Project No. 3.825.2014/K).
We thank the undergraduate students of Mount Allison University
and The George Washington University for their assistance.
\end{acknowledgments}


\begin{thebibliography}{99}
\bibitem{pi0p} J.~Steinberger, W.~K.~H.~Panofsky, and J.~Steller,
 Phys.\ Rev.\ \textbf{78}, 802 (1950);
 A.~Silverman and M.~Stearns, \textit{ibid.} \textbf{83}, 853 (1951);
 W.~K.~H.Panofsky, J.~Steinberger, and J.~Steller,
 \textit{ibid.} \textbf{86}, 180 (1952);
 A.~Silverman and M.~Stearnst, \textit{ibid.} \textbf{88}, 1225 (1952);
 G.~Cocconi and A.~Silverman, \textit{ibid.} \textbf{88}, 1230 (1952).

\bibitem{pi+n} E.~M.~McMillan, J.~M.~Peterson and R.~S.~White,
 Science \textbf{110}, 579 (1949);
 J.~Steinberger and A.~S.~Bishop, Phys.\ Rev.\ \textbf{78}, 493 (1950);
 \textbf{78}, 494 (1950);
 A.~S.~Bishop, J.~Steinberger, and L.~J.~Coom,
 \textit{ibid.} \textbf{80}, 291 (1950);
 J.~Steinberger and A.~S.~Bishop, \textit{ibid.} \textbf{86}, 171 (1952).

\bibitem{pi-p} R.~S.~White, M.~J.~Jacobson, and A.~G.~Schulz,
  Phys.\ Rev.\ \textbf{88}, 836 (1952). 
\bibitem{Fermi}H.~L.~Anderson, E.~Fermi, R.~Martin, and D.~E.~Nagle, Phys.\
	Rev.\ \textbf{91}, 155 (1953).
\bibitem{du13} M.~Dugger \textit{et al.},
  Phys.\ Rev.\ C\ \textbf{88}, 065203 (2013).
\bibitem{PDG} K.~A. Olive  \textit{et al.}, (Particle Data Group),
  Chin.\ Phys.\ C \textbf{38}, 090001 (2014).
\bibitem{Bartalini_05} O.~Bartalini \textit{et al.},
  Eur.\ Phys.\ J.\ A\ \textbf{26}, 399 (2005).
\bibitem{Thiel_12} A.~Thiel \textit{et al.},
  Phys.\ Rev.\ Lett.\ \textbf{109}, 102001 (2012).
\bibitem{Hartmann_14} J.~Hartmann \textit{et al.},
  Phys.\ Rev.\ Lett.\ \textbf{113}, 062001 (2014).
\bibitem{Gottschall_14} M.~Gottschall \textit{et al.},
  Phys.\ Rev.\ Lett.\ \textbf{112}, 012003 (2014).
\bibitem{Sikora_14} M.~H.~Sikora \textit{et al.},
  Phys.\ Rev.\ Lett.\ \textbf{112}, 022501 (2014).
\bibitem{CB}A.~Starostin \textit{et al.}, Phys.\ Rev.\ C\
		\textbf{64}, 055205 (2001).
\bibitem{TAPS} R.~Novotny,
               IEEE Trans.\ Nucl.\ Sci.\ {\bf 38}, 379 (1991).
\bibitem{TAPS2} A.~R.~Gabler \textit{et al.}, Nucl.\ Instrum.\
                Methods\ Phys.\ Res.\ A\ \textbf{346}, 168 (1994).
\bibitem{MAMI} H.~Herminghaus {\it et al.},
         IEEE Trans.\ Nucl.\ Sci.\ {\bf 30}, 3274 (1983).
\bibitem{MAMIC} K.-H.~Kaiser \textit{et al.}, Nucl.\ Instrum.\
                Methods\ Phys.\ Res.\ A\ \textbf{593}, 159 (2008).
\bibitem{slopemamic} S.~Prakhov {\it et al.},
                 Phys.\ Rev.\ C\ {\bf 79}, 035204 (2009).
\bibitem{TAGGER} I.~Anthony {\it et al.},
  Nucl.\ Instrum.\ Methods\ Phys.\ Res.\ A\ {\bf 301}, 230 (1991).
\bibitem{TAGGER1} S.~J.~Hall {\it et al.},
  Nucl.\ Instrum.\ Methods\ Phys.\ Res.\ A\ {\bf 368}, 698 (1996).
\bibitem{TAGGER2} J.~C.~McGeorge {\it et al.},
  Eur.\ Phys.\ J.\ A\ {\bf 37}, 129 (2008).
\bibitem{EtaMassA2} A.~Nikolaev \textit{et al.},
  Eur.\ Phys.\ J.\ A\ \textbf{50}, 58 (2014).
\bibitem{etamamic} E.~F.~McNicoll \textit{et al.},
                 Phys.\ Rev.\ C\ \textbf{82}, 035208 (2010).
\bibitem{SAID} W.~J.~Briscoe, D.~Schott, I.~I.~Strakovsky, and R.~L.~Workman,
 Institute of Nuclear Studies of The George Washington University Database:
	http://gwdac.phys.gwu.edu/analysis/pr\_analysis.html .
\bibitem{cm12} R.~L.~Workman, M.~W.~Paris, W.~J.~Briscoe, and I.~I.~Strakovsky,
	Phys.\ Rev.\ C\ \textbf{86}, 015202 (2012).
\bibitem{hornidge13} D.~Hornidge \textit{et al.},
	Phys.\ Rev.\ Lett.\ \textbf{111}, 062004 (2013).
\bibitem{be06} R.~Beck, R.~Leukel and A.~Schmidt, 
  Acta\ Phys.\ Pol.\ B\ \textbf{33}, 813 (2002); 
 R.~Beck, Eur.\ Phys.\ J.\ A\ \textbf{28}, s01, 173 (2006).
\bibitem{du07} M.~Dugger \textit{et al.},
  Phys.\ Rev.\ C\ \textbf{76}, 025211 (2007).
\bibitem{ba05} O.~Bartholomy \textit{et al.},
  Phys.\ Rev.\ Lett.\ \textbf{94}, 012003 (2005);
  H.~van~Pee \textit{et al.}, Eur.\ Phys.\ J.\ A\ \textbf{31}, 61 (2007).
\bibitem{cr11} V.~Crede \textit{et al.},
       Phys.\ Rev.\ C\ \textbf{84}, 055203 (2011).
\bibitem{maid} The MAID analyses are available through the Mainz website:
	http://wwwkph.kph.uni-mainz.de/MAID/; see also D.~Drechsel, S.~S.~Kamalov,
	and L.~Tiator, Eur.\ Phys.\ J.\ A\ \textbf{34}, 69 (2007).
\bibitem{bnga} The Bonn-Gatchina analyses are available through the Bonn website:
	http://pwa.hiskp.uni-bonn.de/; see also E.~Gutz \textit{et al.},
	Eur.\ Phys.\ J.\ A\ \textbf{50}, 74 (2014).
\bibitem{arndt02} R.~A.~Arndt, W.~J.~Briscoe, I.~I.~Strakovsky,
    and R.~L.~Workman, Phys.\ Rev.\ C\ \textbf{66}, 055213 (2002).
\bibitem{SAID_WI08} R.~L.~Workman, R.~A.~Arndt, W.~J.~Briscoe, M.~W.~Paris,
    and I.~I.~Strakovsky, Phys.\ Rev.\ C\ \textbf{86}, 035202 (2012).
\bibitem{Dieterle_14} M.~Dieterle \textit{et al.},
  Phys.\ Rev.\ Lett.\ \textbf{112}, 142001 (2014).
\bibitem{Althoff_79} K.~H.~Althoff \textit{et al.},
  Z.\ Phys.\ C:\ Particles and Fields \textbf{1}, 327 (1979).
\end{thebibliography}
\end{document}